\documentclass[%
 reprint,
 amsmath,amssymb,
 aps,
]{revtex4-2}
\usepackage{tikz}
\usepackage{lipsum}
\usepackage{graphicx}
\usepackage{dcolumn}
\usepackage{bm}
\usepackage{subcaption}

\begin{document}

\preprint{APS/123-QED}

\title{Deflection of Massive Spin-$\frac{1}{2}$ Particles around Kerr Black Hole}

\author{Haida Li}
\email{eqwaplay@scut.edu.cn}
\affiliation{School of Physics and Optoelectronics, South China University of Technology, Guangzhou 510641, China}

\author{Xiangdong Zhang} 
\email{Corresponding author: scxdzhang@scut.edu.cn}
\affiliation{School of Physics and Optoelectronics, South China University of Technology, Guangzhou 510641, China}

\begin{abstract}
The exact measurement of neutrino mass remains a longstanding issue. So far, there has been much success in providing an upper bound for the neutrino rest mass, both theoretically and experimentally. In this work, by exploring the critical radius of a beam of polarized quantum spin-$\frac{1}{2}$ particle deflecting around a classical Kerr black hole, we attempt to provide an additional testing ground for neutrino mass, as well as the mass of other proposed ultra-light particles yet to be determined. Notably, the quantum Dirac equation is used to derive a MPD-like equation satisfied by the polarized beam of massive spin-$\frac{1}{2}$ particles and identify the effective spin in the spin tensor with the particle's intrinsic quantum spin, confirming the previous theoretical result that the MPD equation can be in fact applied to particles' intrinsic spin. The result of this work shows that corrections of relative magnitude $>10^{-12}$ can be achieved for spin-$\frac{1}{2}$ particles with rest mass equal to $1 eV/c^2$ deflecting around a solar mass Kerr black hole. Although highly theoretical, a new method of extracting the lower bound for the neutrino mass individually is also proposed due to the behavior of the quantum spin correction.
\end{abstract}

\maketitle


\section{\label{sec:111}Introduction}

The neutrino mass problem \cite{Wolfenstein:1977ue,Peccei:1977hh,Mikheev:1986wj,Planck:2018vyg,DESI:2019jxc} is a fundamental challenge in modern particle physics, centered on the fact that neutrinos have mass, a property not predicted by the original Standard Model of particle physics. The problem lies with how neutrinos acquire mass. Nonzero neutrino masses are not possible without the existence of new fundamental fields, beyond those that are part of the standard model. Therefore, the exact measurement of neutrino mass might give key insight into the physics beyond the standard model of particle physics. Currently, the KATRIN experiment \cite{KATRIN:2021uub,KATRIN:2024cdt} achieves the most accurate measurement of neutrino mass, which measures the upper bound of the mass of the electron neutrino at $<0.45$ eV. For the lower bound of neutrino mass, the results from neutrino oscillation experiments \cite{Formaggio:2021nfz} indicate that at least one neutrino mass must be greater than approximately 0.05 eV, and the sum of all three neutrino masses is greater than about 0.06 eV for the normal ordering, or 0.10 eV for the inverted ordering. However, so far, there has been no direct measurement of the lower bound of the mass of the lightest neutrino separately.

In this work, by computing the corrections of the intrinsic quantum spin to the trajectory of a beam of polarized massive spin-$\frac{1}{2}$ particles (e.g. neutrinos) passing by a central Kerr black hole, not only can we directly obtain relatively large quantum corrections versus particles of the same rest mass but without spin, individual lower bounds of the rest mass of the particle can be determined so long as a full ring of critical radius around the black hole can be measured.

For massive particles that have non-zero intrinsic spin, we focus on whether or not there is an extra deflection of the trajectory of quantum particles by the black hole as a result of the particle's intrinsic spin. Classically, the movement of the rotating body is governed by the MPD equations \cite{Mathisson:1937zz,Papapetrou:1951pa,Dixon:1974xoz}. Meanwhile, although the spin angular momentum of a classical spinning body is inherently different from the intrinsic spin of quantum particles, there have been, in fact, some works suggesting the possibility for quantum particles with intrinsic spin to also follow MPD-like equations as the leading order contribution to the solution of quantum Dirac equations \cite{Rudiger:1981uu,Audretsch:1981wf,Oancea:2022utx,Battista:2022vvl}. By combining these results, the quantum spin correction of the particle's trajectory passing by the central black hole can be explicitly computed.

The gravitational lensing effect of massive particles with non-zero spin has already been investigated perturbatively in \cite{Zhang:2022rnn} in the weak field limit. Meanwhile, the particle motion around Kerr black hole has also been investigated previously, although not focusing on the critical radius, by some works \cite{Zhang:2018ocv,Liu:2019wvp,Wu:2021pgf}. As a direct astronomically observable quantity, the critical radius has been theorized as a tool for testing the effect of quantum gravity, e.g., in \cite{Li:2024afr,Li:2025zdt}. 

There are several key results we achieved in this work: First, the deflection of massive spin-$\frac{1}{2}$ particles near the black hole is computed by focusing on the critical radius, namely the innermost possible distance any test particle can achieve during its entire trajectory without being absorbed into the black hole. Second, although the lensing effect of massive particles with non-zero spin has been studied by solving the MPD equations, e.g., in \cite{Zhang:2022rnn}, it is assumed that quantum particles, e.g. electrons, neutrinos, etc., will follow a similar trajectory as given by the MPD equations. In this work, by combining previous results in \cite{Rudiger:1981uu,Audretsch:1981wf} and calculations to specifically identify the effective spin provided by the quantum spin tensor given in \cite{Rudiger:1981uu}, we can show that quantum particles, particularly highly polarized particle beams, indeed exhibit effective quantum spin corrections to their trajectory passing by classical black holes. Additionally, the exact corrections of particles with given rest mass while traveling with a speed of $v$ around a black hole with given mass is computed and analyzed, the results show that the correction may not be negligible when both the particles rest mass and black hole mass are small. Finally, although still highly theoretical, a potential new method of extracting the lower bound for the neutrino mass is proposed based on the computed results. The relatively high level of correction and our newly proposed way to obtain a lower bound for the test particle's rest mass might enable new possibilities of gravitational lensing/deflection-related experiments in both high-precision astronomy and microscopic tests, e.g., analog gravity experiments \cite{Schutzhold:2002rf,Steinhauer:2015saa,Chakraborty:2022zfl}.

The structure of this work is as follows. In Section \ref{sec2}, we first introduce the MPD-like equation that is satisfied by the quantum particle and identify the effective spin generated by the quantum spin tensor. Then, we also briefly introduce the theoretical background of the classical computation of the trajectory of a spinning body around a black hole. In Section \ref{sec3}, several key results will be presented. First, we will show that whether the particle is relativistic does not significantly affect the quantum spin correction, suggesting that the magnitude of the correction mainly comes from the change in the particle's rest mass, as well as the size of the central black hole. Then, we will analyze in detail the dependencies of the quantum spin correction to the particle's rest mass, the size of the center black hole, as well as how fast the black hole is rotating. Also, based on the results computed, we propose a new method of extracting the lower bound for the mass of an ultralight particle with non-zero spin, potentially include neutrinos. Finally, we will provide some discussion on the results we obtained in this work in Section \ref{sec4}.

\section{Theoretical Background}\label{sec2}

\subsection{The effect of quantum spin on particle trajectory}

First, recall that, given a spacetime with the metric $g_{\mu\nu}$ and the corresponding Riemann tensor $R^\mu{}_{\nu \rho \sigma}$, the equations of motion of a classical spinning object can be described by the Mathisson-Papapetrou-Dixon (MPD) equations \cite{Mathisson:1937zz,Papapetrou:1951pa,Dixon:1974xoz} as:
\begin{equation}\label{mpd1}
\begin{aligned}
&\frac{\mathrm{D} p^\mu}{\mathrm{d} \tau} =-\frac{1}{2} R^\mu{}_{\nu \rho \sigma} u^\nu S^{\rho \sigma} \\
&\frac{\mathrm{D} S^{\mu \nu}}{\mathrm{d} \tau} =2 p^{[\mu} u^{\nu]},
\end{aligned}
\end{equation}
where $p^{\mu}$ is the generalized four-momentum, $u^{\mu}$ is the four-velocity, $S^{\mu\nu}$ is the antisymmetric spin tensor that characterizes the test particle's spin angular momentum, and for an arbitrary vector field $T^{\mu}$, the derivative $\frac{D}{d\tau}$ is defined as:
\begin{equation}
\frac{\mathrm{D} T^\mu}{\mathrm{d} \tau} \equiv \frac{\mathrm{~d} T^\mu}{\mathrm{d} \tau}+\Gamma_{\rho \sigma}^\mu T^\rho u^\sigma.
\end{equation}
The above equations contain fewer constraints than independent variables and thus, to solve the MPD equations, additional conditions must be imposed. A common choice is the Tulczyjew-Dixon constraint \cite{tulczyjew1959motion}:
\begin{equation}\label{const1}
S^{\mu \nu} p_\nu=0.
\end{equation}
Under this constraint, $S^{\mu\nu}$ and $p^{\mu}$ satisfy the corresponding normalization conditions \cite{Zalaquett:2013ifd}:
\begin{equation}\label{eqn111}
\begin{aligned}
p^\mu p_\mu & =-m^2 & (m \geq 0) \\
\frac{1}{2} S^{\mu \nu} S_{\mu \nu} & =J_m^2 & \left(J_m \geq 0\right),
\end{aligned}
\end{equation}
where $m$ is the rest mass of the test particle and $J_m$ is the size of its spin angular momentum.

Now, note that so far we have only introduced the equations of motion satisfied by classical objects. The macroscopic spin angular momentum mainly comes from, instead of the quantum spin of the constituting particles that form the classical object, the cumulative effects of the orbital angular momenta and EM forces between the microscopic particles that comprise the macroscopic object. The classical trajectory of a spinning object described by the MPD equations is thus completely different from the quantum spin, which is an intrinsic property of quantum particles.

However, the possibility of effective effects coming from intrinsic quantum spin has been discussed previously \cite{Rudiger:1981uu,Audretsch:1981wf} by considering WKB expansions of spinning massive particles as an approximation of the solution to the Dirac equation in curved space-time:
\begin{equation}
\Psi=e^{-i S / \hbar}\left(\Psi^{(0)}+\hbar \Psi^{(1)}+\cdots\right),
\end{equation}
where $S$ is a phase parameter satisfying $\partial_\mu S \partial^\mu S=m^2$ by the requirement of the Dirac equation. It is explicitly shown in \cite{Rudiger:1981uu} that up to the first order in $\hbar$, for massive particles with spin-$\frac{1}{2}$, the quantum-corrected equation of motion of $\Psi$ resembles (\ref{mpd1}) by making the following identification:
\begin{equation}\label{spinq}
\begin{split}
    S^{\mu \nu}&=\frac{1}{2} \hbar(\mathrm{i} \bar{\Psi} \Psi)^{-1} \bar{\Psi} \sigma^{a b} \Psi\\
    &=\frac{1}{2} \hbar\left(\mathrm{i} \bar{\psi}^{(0)} \psi^{(0)}\right)^{-1} \bar{\psi}^{(0)} \sigma^{\mu \nu} \psi^{(0)}+\ldots.
\end{split}
\end{equation}
where the original paper used the convention $\sigma^{\mu\nu}:=\gamma^{[\mu}\gamma^{\nu]}$ such that both $i\bar{\Psi}\Psi$ and $\bar{\Psi}\sigma^{\mu\nu}\Psi$ are real numbers. We will keep the same convention here.

When we consider a beam of polarized particles, the leading order quantum effect on the particle trajectory, under the influence of a classical macroscopic gravitational field, can thus be approximately described by substituting eqn. (\ref{spinq}) in eqn. (\ref{mpd1}).

However, it is in fact previously unknown whether the "effective" spin tensor $S^{\mu\nu}$ in eqn. (\ref{spinq}) actually satisfies eqn. (\ref{eqn111}) and how it will actually affect the trajectory of the particles. To show that eqn. (\ref{eqn111}) is indeed satisfied, we now compute:
\begin{equation}\label{spinqq}
\begin{split}
    S^{\mu \nu}S_{\mu \nu}=\frac{1}{4} \hbar^2\left(\mathrm{i} \bar{\psi}^{(0)} \psi^{(0)}\right)^{-2} \bar{\psi}^{(0)} \sigma^{\mu \nu} \psi^{(0)}\bar{\psi}^{(0)} \sigma_{\mu \nu} \psi^{(0)}.
\end{split}
\end{equation}

For an arbitrary spacetime, given a beam of massive particles whose spin is polarized along the direction $\vec{n}$, we will first compute this term in the particle's rest frame and then transform it to the frame of an inertial observer. The normalized spinor for the particle is:
\begin{equation}
u=\sqrt{2 m}\binom{\chi}{0}, \quad \bar{u}=u^{\dagger} \gamma^0=\sqrt{2 m}\left(\begin{array}{ll}
\chi^{\dagger} & 0
\end{array}\right),
\end{equation}
where $\chi$ is a 2-component spinor satisfying $\chi^{\dagger} \chi=-i$. Then:
\begin{equation}
i\bar{u} u=2 m,
\end{equation}
and by using the properties of the gamma matrices, we have:
\begin{equation}
\begin{split}
    &\bar{u} \sigma^{i j} u=2 m \epsilon^{i j k} n_k, \quad i,j,k=1,2,3\\
    &\bar{u} \sigma^{0i} u=\bar{u} \sigma^{00} u=0,
\end{split}
\end{equation}
which leads to:
\begin{equation}\label{spinqq2}
\begin{split}
    S^{\mu \nu}S_{\mu \nu}&=\frac{1}{4} \hbar^2\left(\mathrm{i} \bar{\psi}^{(0)} \psi^{(0)}\right)^{-2} \bar{\psi}^{(0)} \sigma^{\mu \nu} \psi^{(0)}\bar{\psi}^{(0)} \sigma_{\mu \nu} \psi^{(0)}.\\
    &=\frac{1}{4} \hbar^2 (2m)^{-2}(2m)^2 (\epsilon^{i j k} n_k)(\epsilon_{i j l} n^l)\\
    &=\frac{\hbar^2}{2} n^kn_k=\frac{\hbar^2}{2} .
\end{split}
\end{equation}
The final result is a number that only depends on the particle spin and is thus invariant in any reference frame. Compare this result to eqn (\ref{eqn111}), it is straightforward to see that not only does the effective spin satisfy all of the properties of the classical spin tensor in MPD equations, but also $J_m=\frac{\hbar}{2}$, suggesting that the trajectory of a beam of polarized spin-$\frac{1}{2}$ massive particles can indeed be described by the effective MPD equations up to $\hbar$ order.

\subsection{The critical radius of polarized spin massive quantum particles}

In this paper, we focus on the axisymmetric spacetime with the following metric:
\begin{equation}\label{generalG}
\mathrm{d} s^2=-A \mathrm{~d} t^2+B \mathrm{~d} t \mathrm{~d} \varphi+D \mathrm{~d} r^2+C \mathrm{~d} \varphi^2+F \mathrm{~d} \theta^2,
\end{equation}
where $x^\mu=(t, r, \theta, \varphi)$ are the coordinates and A, B, C, D, F are the functions of $r$ and $\theta$ only. Specifically, the equatorial plane of the Kerr spacetime can be obtained by setting $\theta=\frac{\pi}{2}$ \cite{creighton1973cw}:
\begin{equation}
\begin{aligned}
&A(r)=1-\frac{2 M}{r}, \quad B(r)=-\frac{4 a M}{r},\quad F=0 \\
&C(r)=r^2+a^2+\frac{2 M a^2}{r},\quad D(r)=\frac{r^2}{r^2-2 M r+a^2},\\
\end{aligned}
\end{equation}
where $M$ is the mass of the central black hole, and $a=\frac{J_{K}}{M}$ is the angular momentum of the black hole per unit mass. Additionally, this metric reduces to the Schwarzschild spacetime when the rotation parameter $ a$ is set to zero.

We focus on the test particle's critical radius, namely the innermost radius achievable by particles passing through the center black hole with arbitrary impact parameter $b$. This radius provides us with the most direct observable related to the strong black hole gravitational lensing effect occurring in the vicinity of the central black hole.

The calculations of the critical radius of spinning massive particles in static spherically symmetric spacetimes are quite straightforward. (also for the innermost stable orbit of stationary axisymmetric spacetimes in the equatorial plane, see \cite{Duan:2023gvm}) First, for particles satisfying eqn. (\ref{mpd1}), we consider the case when it's vertical spin:
\begin{equation}
S_\mu=\frac{\sqrt{-g}}{2 m} \varepsilon_{\mu \alpha \beta \gamma} S^{\alpha \beta} p^\gamma,
\end{equation}
is perpendicular to the equatorial plane, i.e., $S^t=S^r=S^{\varphi}=0, S^\theta \neq 0$, where $g=\mathrm{det}(g_{\mu\nu})$. In order for the constraint (\ref{const1}) to be satisfied under this setting, we have $p^{\theta}=0$. A solution can be obtained by setting:
\begin{equation}
p^\theta=0=S^{\mu \theta}
\end{equation}
We further assume that:
\begin{equation}
\partial_\theta A=\partial_\theta B=\partial_\theta C=\partial_\theta D=0,
\end{equation}
which is automatically satisfied on the equatorial plane of Kerr spacetime. Under these settings, the following relation can be obtained \cite{Zhang:2022rnn}:

\begin{widetext}
\begin{equation}\label{rdot}
\begin{aligned}
\dot{r}=\eta_3^2\left(B^2+4 A C\right) \sqrt{D\left[64\left(-A \eta_1^2+B \eta_1 \eta_2+C \eta_2^2\right)\left(B^2+4 A C\right)^{-1} \eta_3^{-2}-m^2\right]},
\end{aligned}
\end{equation}
\end{widetext}
where:
\begin{equation}\label{eta1}
\begin{aligned}
& \eta_1=4 L+\alpha\left(B^{\prime} L+2 C^{\prime} E\right),\\
&\eta_2=4 E+\alpha\left(2 A^{\prime} L-B^{\prime} E\right),\\
&\eta_3=-16+\alpha^2\left(B^{\prime 2}+4 A^{\prime} C^{\prime}\right) ,
\end{aligned}
\end{equation}
where $\alpha$ is defined as:
\begin{equation}
\alpha=-\frac{s_j j}{\sqrt{D\left(B^2 / 4+A C\right)}},
\end{equation}
and:
\begin{equation}
j=\frac{J_m}{m}
\end{equation}
is the particles spin to mass ratio.

The constants of motion $E$ and $L$ can be written in terms of the impact factor $b$ as:
\begin{equation}\label{EL}
\begin{aligned}
& E=\left.p^t\right|_{r \rightarrow \infty}=\frac{m}{\sqrt{1-v^2}}, \\
& L=\left.\left(r^2 p^{\varphi}+s_j j p^t\right)\right|_{r \rightarrow \infty}=\frac{s_o m v b}{\sqrt{1-v^2}}+\frac{s_j j m}{\sqrt{1-v^2}} .
\end{aligned}
\end{equation}
where $m$ is the test particle's rest mass. The particle's closest distance $r_0$ to the black hole during its trajectory satisfies the following equation:
\begin{equation}\label{eqnr}
\left.\dot{r}\right|_{r=r_0}=0,
\end{equation}
Using eqn. (\ref{rdot}), (\ref{eta1}), and (\ref{EL}), the impact factor $b$ can be solved from eqn. (\ref{eqnr}).

Additionally, let $r_c$ be the critical radius of the test particle. The following equation will be satisfied if a particle with critical impact parameter $b_c$ can exactly reach its critical radius:
\begin{equation}\label{eqI}
\left.\frac{\mathrm{d} \dot{r}^2}{\mathrm{~d} r}\right|_{r=r_c}=0.
\end{equation}
The critical radius of massive particles with non-zero spin in arbitrary spacetime with the metric described by eqn (\ref{generalG}) can thus be computed by solving eqn. (\ref{eqI}). 

\section{Main Results}\label{sec3}

In this work, we mainly focus on computing the quantum spin corrections of the critical radius of the trajectory of a massive particle beam with polarized spin passing by a Kerr black hole. Most importantly, by analyzing the numerically obtained critical radius, we will show that, not only does the quantum spin of the polarized particle beam provide noticeable corrections to the particle's critical radius, but it also generates a unique lower bound for the particle's mass, if a complete ring of critical radius can be observed without intersecting with the black hole's horizon.

\begin{figure}[h!]
\includegraphics[height=5cm]{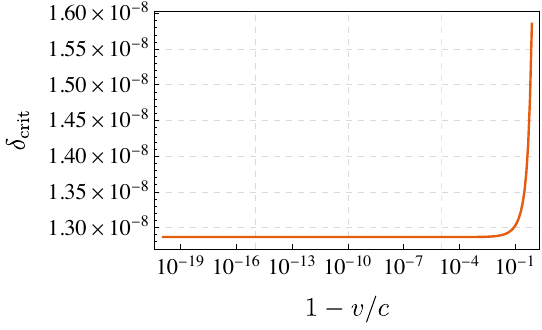} 
\caption{\label{fig:1} Relation between particle speed $v$ and the spin correction to the critical radius.}
\end{figure}

\begin{figure*}[h!tb]
    \centering
    \begin{subfigure}[b]{0.48\textwidth}
        \centering
        \includegraphics[height=4.5cm]{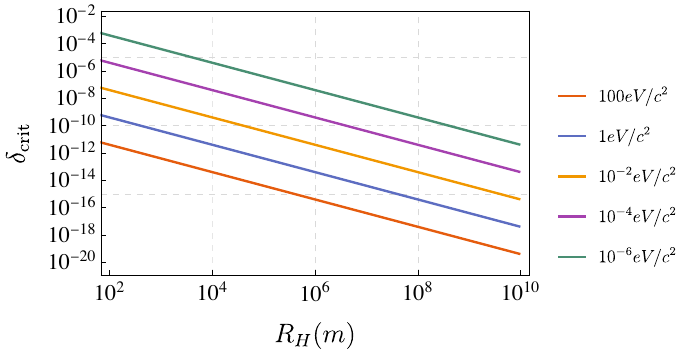} 
        \caption{}
        \label{fig:sub1}
    \end{subfigure}
    \hfill
    \begin{subfigure}[b]{0.48\textwidth}
        \centering
        \includegraphics[height=5cm]{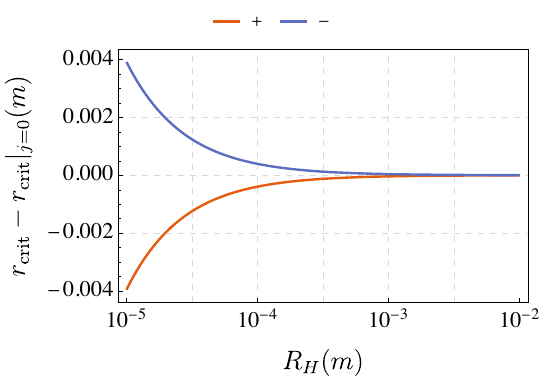} 
        \caption{}
        \label{fig:sub2}
    \end{subfigure}
    \hfill
    \begin{subfigure}[b]{0.48\textwidth}
        \centering
        \includegraphics[height=4.5cm]{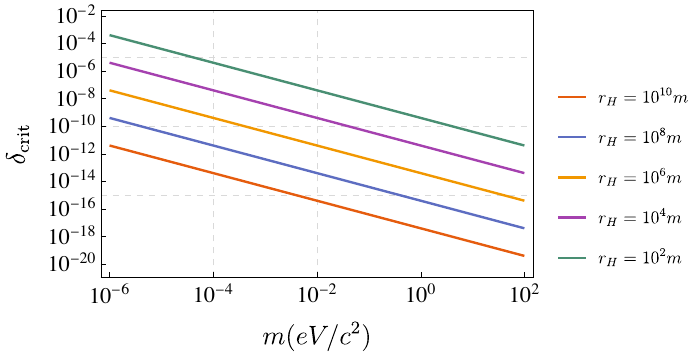} 
        \caption{}
        \label{fig:sub3}
    \end{subfigure}
    \hfill
    \begin{subfigure}[b]{0.48\textwidth}
        \centering
        \includegraphics[height=5cm]{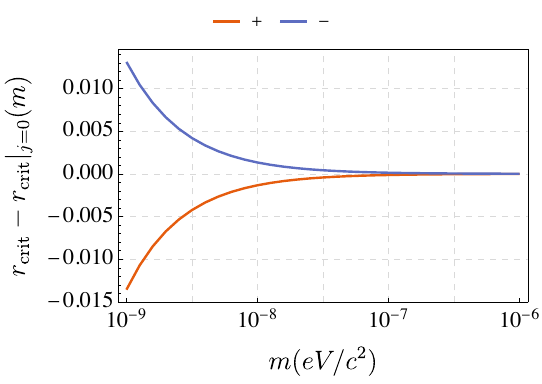} 
        \caption{}
        \label{fig:sub4}
    \end{subfigure}
    \hfill
    \begin{subfigure}[b]{0.48\textwidth}
        \centering
        \includegraphics[height=4.5cm]{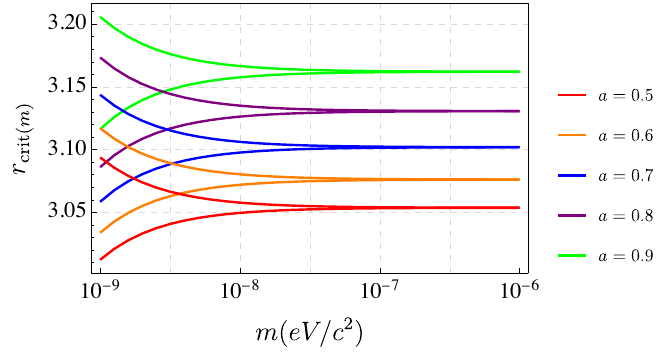} 
        \caption{}
        \label{fig:sub5}
    \end{subfigure}
    \hfill
    \begin{subfigure}[b]{0.48\textwidth}
        \centering
        \includegraphics[height=4.6cm]{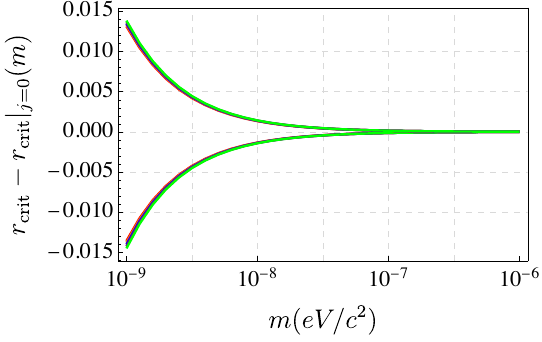} 
        \caption{}
        \label{fig:sub6}
    \end{subfigure}
\caption{The main results of our work: (a) The relative correction of the polarized particle spin to the critical radius of the particle trajectory with respect to the size of the black hole horizon when fixing particle mass. (b) Detailed comparison showing when the radius of the black becomes small, the quantum spin correction to the particle trajectory become large, and the direction of the correction takes opposite sign when the particle spin shares the same direction with its orbital angular momentum or opposite the direction of its orbital angular momentum. (c) The relative correction of the polarized particle spin to the critical radius of the particle trajectory with respect to particle mass when fixing the size of the black hole horizon. (d) Detailed comparison showing when the radius of the black becomes small, the quantum spin correction to the particle trajectory become large, and the direction of the correction takes opposite sign when the particle spin shares the same direction with its orbital angular momentum or opposite the direction of its orbital angular momentum. (e) Comparison between the quantum spin corrections of particles passing through black holes with different rotation parameters. (f) Comparison between the relative difference induced by the quantum spin corrections of particles passing through black holes with different rotation parameters.}
\label{fig:2}
\end{figure*}

First, Fig. \ref{fig:1} shows the relation between the massive particle's speed $v$ and the spin correction to the critical radius. The graph is plotted by choosing a solar mass Schwarzschild black hole with $a=\frac{1}{2}$. The particle spin is fixed to be $J_m=\frac{\hbar}{2}$ and the particle's rest mass is $10^{-3} eV/c^2$. Also, for the rest of this work, we will continue to fix the particle spin as $J_m=\frac{\hbar}{2}$ and only change the particle's rest mass as well as the black hole size. As can be seen in this graph, the correction quickly converges as the particle becomes relativistic, i.e., $v>0.99 c$, suggesting that all ultra-relativistic particles of the same spin-mass ratio and velocity share the same spin correction. This result can be explained by looking at eqn. (\ref{EL}), where it is straightforward to see that for both constants of motion $E$ and $L$, including the spin-related term in the particle's angular momentum, share the same relativistic boost. Thus, as the particle becomes ultra-relativistic, both constants of motion become very large simultaneously. While these large constants of motion become the dominant contributing factor in the quantum spin correction, due to their shared proportionality to the same overall boost term, further increase in particle velocity will not change the quantum spin corrections further, thus explaining the convergence. This can be best illustrated by examining the equation. (\ref{rdot}), where the contributions of $E$ and $L$ are contained homogeneously in the square root.

Fig. \ref{fig:2} shows the quantum spin corrections to the particle's critical radius under various circumstances. First, in Fig. \ref{fig:sub1}, we show how the relative difference of $r_{crit}$ changes with respect to the radius $r_H$ of the horizon of the central black hole, where the relative difference $\delta_{crit}$ is defined as:
\begin{equation}
    \delta_{\mathrm{crit}}:= \bigg|\frac{r_{\mathrm{crit}}-r_{\mathrm{crit}}|_{j=0}}{r_{\mathrm{crit}}|_{j=0}}\bigg|,
\end{equation}
where $r_{\mathrm{crit}}|_{j=0}$ is the critical radius of the particle obtained by only setting $j=0$ and keeping the other parameters the same. From this graph, we can see that, when the rest mass of the particle is kept constant as the black hole size decreases, there is a significant increase in the magnitude of spin correction in terms of the critical radius of the test particles.

In Fig. \ref{fig:sub2}, it is shown that, for a beam of polarized spin-$\frac{1}{2}$ particles with rest mass $m=1 eV/c^2$, using the absolute difference between $r_{\mathrm{crit}}$ and $r_{\mathrm{crit}}|_{j=0}$, when the direction of the spin polarization changes from upwardly perpendicular to the equatorial plane (marked as "+", since we choose both $s_0=1$ and $s_j=1$) to downwardly perpendicular to the equatorial plane (marked as "-", in which case $s_0=1$ and $s_j=-1$), the quantum spin correction takes the opposite sign. That is, when the particle spin has the same direction as its orbital angular momentum moving around the black hole (when only considering movements within the equatorial plane), the quantum spin correction exhibits a centripetal effect on the particle trajectory, while as the particle spin takes the opposite direction against its orbital angular momentum, the quantum spin correction exhibits a centrifugal effect.

For the impact of particle's rest mass, effects similar to Fig. \ref{fig:sub1} and Fig. \ref{fig:sub2} are shown in Fig. \ref{fig:sub3} and Fig. \ref{fig:sub4}: When keeping the black hole radius constant, as the particle mass decreases, there is a significant increase in the magnitude of spin correction in terms of the critical radius of the test particles. Noticeably, corrections of relative magnitude $>10^{-12}$ can be achieved for spin-$\frac{1}{2}$ particles with rest mass equal to $1 eV/c^2$ deflecting around a solar mass ($r_H\sim 3\times10^3 m$) Kerr black hole. This result further confirms that both particle energy (effecting its spin/mass ratio) and the black hole radius have an impact on the quantum spin correction on the trajectory of spin massive particles. Also, in Fig. \ref{fig:sub4}, the same centripetal/centrifugal effect for $+$/$-$ signs is observed for a solar-mass black hole with $a=\frac{1}{2}$.

Interestingly, from Fig. \ref{fig:sub2} and Fig. \ref{fig:sub4}, we can also see another phenomenon: when fixing the black hole radius, there is a lower limit of particle mass at which, on one side, the critical radius falls directly into the black horizon. This lower bound arises because for masses below a critical value $m_{\text{crit}}$, the spin correction pulls the critical radius inside the event horizon, making its observation impossible. Therefore, the observation of a whole critical ring implies $m > m_{\text{crit}}$. 

Under ideal circumstances (i.e. when the black hole is small enough and the neutrino beam is perfectly polarized), this effect can be used to produce an experiment that can give a unique lower limit to the mass of ultra-light massive particles with non-zero spin, including potentially neutrinos: if we can observe a critical radius larger than the black hole horizon, then the particle's rest mass is guaranteed to be larger than $m_{crit}$. Unfortunately, the criterion by which a full ring of critical radius can be observed is quite strict and thus is far beyond the current observational limits. For example, for a solar mass black hole, only an extremely small lower bound of $m\sim10^{-10} eV/c^2$ can be given. Therefore, the practical application of this method relies on either the test particle's mass being extremely small or the black hole being microscopic.

Also, when considering black holes with different rotation speeds, from Fig. \ref{fig:sub5}, we can see that the absolute critical radius is different for each setting of the rotation parameter $a$. Specifically, as $a$ becomes larger, the critical radius becomes larger. Fig. \ref{fig:sub6} shows the comparison between the relative difference induced by the quantum spin corrections of particles passing through black holes with different rotation parameters. As can be seen in this graph, the relative corrections induced by the quantum spin are almost the same for different rotating black holes. This effect can, in fact, be partially explained by looking at results, e.g., like eqn. (5.5) in \cite{Zhang:2022rnn}, where it is shown that the leading order corrections of the black hole rotation to the deflection of the particle trajectory are almost completely separated from the effect of particle spin.

\section{Discussion}\label{sec4}
In summary, there are three key points that we would like to make regarding the results obtained:

Firstly, based on our study in this work, we discovered that an increase in spin correction to the critical radius of the test particle occurs as both the mass of the test particle and the central black hole decrease. Noticeably, corrections of relative magnitude $>10^{-12}$ can be achieved for spin-$\frac{1}{2}$ particles with rest mass equal to $1 eV/c^2$ deflecting around a solar mass ($r_H\sim 3\times10^3 m$) Kerr black hole. Meanwhile, as the direction of the spin polarization changes, particularly when the direction changes from upwardly perpendicular to the equatorial plane to downwardly perpendicular to the equatorial plane, the quantum spin correction takes the opposite sign. Specifically, when the particle spin has the same direction as its orbital angular momentum moving around the black hole, the quantum spin correction exhibits a centripetal effect on the particle trajectory, while when the particle spin takes the opposite direction against its orbital angular momentum, the quantum spin correction exhibits a centrifugal effect. Also, when considering black holes with different rotation speeds, we can observe that while the absolute critical radius is different for each setting of the rotation parameter $a$, the relative corrections induced by the quantum spin are almost the same for different rotating black holes.

Secondly, one interesting result we obtain from this work is that for a given black hole with fixed mass and rotation speed, a lower bound for the particle's rest mass can be given: When the particle spin has the same direction as its orbital angular momentum, the critical radius of the test particle becomes smaller due to the effect of the intrinsic quantum spin, and the magnitude of this effect increases as the particle's spin/mass ratio becomes small. Naturally, for a particle of spin-$\frac{1}{2}$, there is a lower limit on the particle's rest mass, below which the reduced critical radius becomes even smaller than the black hole's horizon radius. In other words, so long as a full ring of critical radius can be observed, the particle's rest mass has to be above a certain value for a given black hole, and this lower limit increases as the size of the given black hole decreases. This phenomenon can be potentially used as a neutrino mass detector to provide a separate test of the lower bound of neutrino mass. Also, there are other proposed ultra-light particles \cite{Poddar:2022qbt,Hubisz:2024hyz,Govindarajan:2024jxc,Eberhardt:2025caq,Bartnick:2025lbv} that can also be detected by employing the methods proposed in this work. Unfortunately, the criterion by which a full ring of critical radius can be observed is quite strict and is far beyond the current observational limits for current measurements of neutrino mass. For example, for a solar mass black hole, only an extremely small lower bound of $m<10^{-10} \mathrm{eV}/c^2$ can be detected. Therefore, the practical application of this method relies on either the test particle's mass being extremely small or the black hole being microscopic.

Thirdly, although we believe that our analysis of the quantum spin corrections via MPD-like effective equations is accurate, at least to the order of $\hbar$, there might still be other factors that need to be taken into account when considering the effect of quantum spin on the particle's trajectory. One theoretical phenomenon that may contribute additionally is the gravitational spin Hall effect (also called spin-gravity coupling effect), according to papers like \cite{Bliokh:2015yhi,Oancea:2019pgm,Wang:2023bmd}, this effect describes a quantum spin correction to the particle's trajectory when the particle is traveling roughly along the same direction as the gravitational force it receives. When considering the specific case of particle beams emitting from a distant source, it is reasonable to believe that, as this effect mainly occurs at a scale much larger than the size of the black hole, the total quantum spin-hall effect as the particle enters and then exits the black hole should be self-canceling, thus making the overall effect minimal. However, there is currently no theory available to compute this effect for relativistic particles without imposing the non-relativistic limit in the Foldy–Wouthuysen transformation \cite{Foldy:1949wa,Foldy:1952ukj,Pryce:1948pf,Tani:1951trl}. Therefore, future study along this direction might provide a precise estimation of this effect. We believe that, by incorporating this effect into the current model, a more accurate calculation of the quantum spin correction to the test particle's trajectory can be achieved.

\section*{Acknowledgements}
This work is supported by the National Natural Science Foundation of China (NSFC) with Grant No.12275087.

\providecommand{\noopsort}[1]{}\providecommand{\singleletter}[1]{#1}%

\end{document}